# Strongly hybridized phonons in one-dimensional van der Waals crystals

Shaoqi Sun[1], Qingyun Lin[2], Yihuan Li[1], Daichi Kozawa[3], Huizhen Wu[1], Shigeo Maruyama[4], Pilkyung Moon[5], Toshikaze Kariyado[3]*, Ryo Kitaura[3]* and Sihan Zhao[1]*

1. School of Physics, Zhejiang Key Laboratory of Micro-Nano Quantum Chips and Quantum Control, and State Key Laboratory of Silicon and Advanced Semiconductor Materials, Zhejiang University, Hangzhou 310058, China
2. Center of Electron Microscopy, School of Materials Science and Engineering, Zhejiang University, Hangzhou 310027, China
3. Research Center for Materials Nanoarchitectonics (MANA), National Institute for Materials Science (NIMS), 1-1 Namiki, Tsukuba 305-0044, Japan
4. Department of Mechanical Engineering, The University of Tokyo, Tokyo 113-8656, Japan
5. Arts and Sciences, NYU Shanghai, Shanghai 200124, China; NYU-ECNU Institute of Physics at NYU Shanghai, Shanghai 200062, China

Corresponding authors:
KARIYADO.Toshikaze@nims.go.jp
KITAURA.Ryo@nims.go.jp
sihanzhao88@zju.edu.cn

**Abstract**

The phenomena of pronounced electron-electron and electron-phonon interactions in one-dimensional (1D) systems are ubiquitous, which are well described by frameworks of Luttinger liquid, Peierls instability and concomitant charge density wave. However, the experimental observation of strongly hybridized phonons in 1D was not demonstrated. Herein we report the first observation of strongly hybridized phonons in 1D condensed matters by using double-walled carbon nanotubes (DWNTs), representative 1D van der Waals crystals, with combining the spectroscopic and microscopic tools as well as the ab initio density functional theory (DFT) calculations. We observe uncharted phonon modes in one commensurate and three incommensurate DWNT crystals, three of which concurrently exhibit strongly-reconstructed electronic band structures. Our DFT calculations for the experimentally observed commensurate DWNT (7,7) @ (12,12) reveal that this new phonon mode originates from a (nearly) degenerate coupling between two transverse acoustic modes (ZA modes) of constituent inner and outer nanotubes having approximately trigonal and pentagonal rotational symmetry along the nanotube circumferences. Such coupling strongly hybridizes the two phonon modes in different shells and leads to the formation of a unique lattice motion featuring evenly distributed vibrational amplitudes over inner and outer nanotubes, distinct from any known phonon modes in 1D systems. All four DWNTs that exhibit the pronounced new phonon modes show small chiral angle twists, closely matched diameter ratios of $\frac{3}{5}$ and decreased frequencies of new phonon modes with increased diameters, all supporting the uncovered coupling mechanism. Our discovery of strongly hybridized phonons in DWNTs open new opportunities for engineering phonons and exploring novel phonon-related phenomena in 1D condensed matters.

Phonons are quantized collective lattice motions extended over the entire crystal [1]. They can strongly interact with a rich body of fundamental particles and excitations, giving rise to a wealth of important physical phenomena and consequences [2,3]. Phonon-phonon interactions were first theoretically studied by Peierls in 1929 [4]. The strong phonon-phonon interactions are directly responsible for the low thermal conductivity in some of the best thermoelectric bulk materials, including PbTe and $Bi_2Te_3$ containing heavy elements [5-8]. They can also lead to strong reconstruction of phonon spectra, recently demonstrated in reconstructed two-dimensional (2D) van der Waals moiré structures containing lighter atoms [9-12]. Although strong electron-electron and electron-phonon interactions are ubiquitous in one-dimension (1D) [13-16], strong phonon-phonon interactions are so far not observed and realized in 1D condensed matters.

Double-walled carbon nanotubes (DWNTs), comprised of two concentric and van der Waal force-coupled single-walled carbon nanotubes (SWNTs), are one of the most ideal platforms to investigate the intriguing coupling effect in 1D. Electronic coupling in incommensurate DWNTs was intensively studied both in a perturbative [17-19] and non-perturbative coupling regime [20,21]. Regarding the phonon coupling, previous experimental studies [22-24] and theoretical models [22-27] mostly focused on the weakly-coupled radial breathing modes (RBMs) in DWNTs where blueshifts of RBMs were observed and interpreted as a result of two coupled mechanical oscillators [22-24].

Herein we report the first observation of strongly hybridized phonons in 1D van der Waals materials. We observe a previously uncharted Raman phonon mode both in one commensurate and three incommensurate DWNT crystals. Our density functional theory (DFT) calculation for a commensurate DWNT (7,7) @ (12, 12) reveals that the new phonon mode originates from a strong hybridization of two nearly-degenerate transverse acoustic modes (ZA modes) in two constituent nanotubes. All the four DWNTs exhibiting new phonon modes are found to have approximate diameter ratios of $\frac{3}{5}$, small twist angles and decreased frequencies with increased diameters.

Our study combines electron diffraction, Rayleigh scattering and resonant Raman spectroscopies as well as DFT calculations [28]. Figure 1(a) shows the Rayleigh spectrum of a representative DWNT (12, 5) @ (22, 4) (inset of Fig. 1(b) and Supplementary Note 2) [35,36] where all the observed optical resonances can be assigned to the original optical transitions from constituent inner and outer nanotubes with moderate energy shifts, indicative of an electronic weak-coupling effect (energy shifts approximately within -200 and + 50 meV) [17,18, 37]. $S(M)_{ii}^{i(o)}$ denotes the i-th optical transition from a semiconducting (metallic) nanotube with the superscripts i (o) standing for inner (outer) nanotube throughout this paper. Figure 1(b) shows the resonant Raman spectrum of the same DWNT in the low frequency range excited at 2.33 eV (blue dashed line in Fig. 1(a)). Two prominent RBM peaks, corresponding to the concerted in-phase (low frequency) and out-of-phase (high frequency) motions of the two walls, are present. We refer to an empiral relation $\omega_{RBM} = \frac{228}{d}\ (nm\ cm^{-1})$ to obtain RBM peak positions for pristine inner and outer SWNTs that are indicated by the black ($\omega_i$) and red ($\omega_o$) dashed lines thoughout this paper where $d$ is the nanotube dimater [38]. The frequencies for the coupled motions are blue-shifted by about 12.9 and 7.7 cm$^{-1}$ when compared with the pristine SWNTs, being consistent with previous reports [22-25].

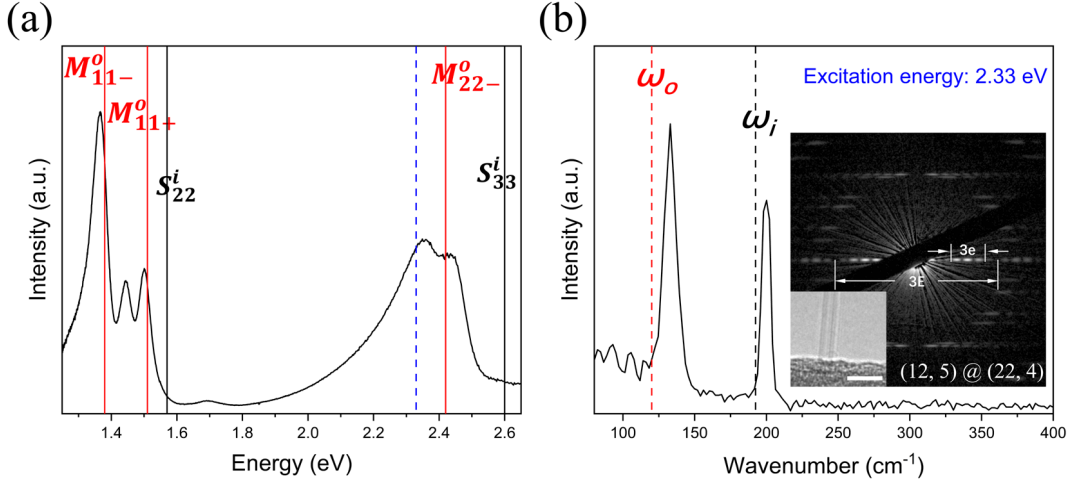

**FIG.1. Characterization of a representative DWNT (12, 5) @ (22, 4).** (a) Rayleigh scattering spectrum. The optical transitions of pristine inner and outer nanotubes are indicated by black and red solid lines, respectively [38]. The superscripts, "i" and "o", represent inner and outer nanotubes, respectively. (b) Resonant Raman spectrum. Excitation energy of ~2.33 eV is indicated by the blue dashed line in (a). In-phase (low frequency) and out-of-phase (high frequency) RBM oscillations are observed. The red and black dashed lines indicate the RBM frequencies of pristine outer ($\omega_o$) and inner ($\omega_i$) nanotubes, respectively. Inset of (b) shows the electron diffraction pattern and the transmission electron microscopy (TEM) image (Scale bar, 5 nm).

We identify a commensurate DWNT with chirality of (7, 7) @ (12, 12) (sample number N = 1) whose electron diffraction and TEM image are presented in Fig. 2(a) [39]. We observe, in Fig. 2(b), a pronounced peak at ~2.05 eV (indicated by the gold frame) that arises from strong inter-tube electronic coupling (i.e., "strong-coupling" [20]). To the best of our knowledge, this is the first experimental confirmation of significant change of electronic band structure in a commensurate DWNT. Figure 2(c)

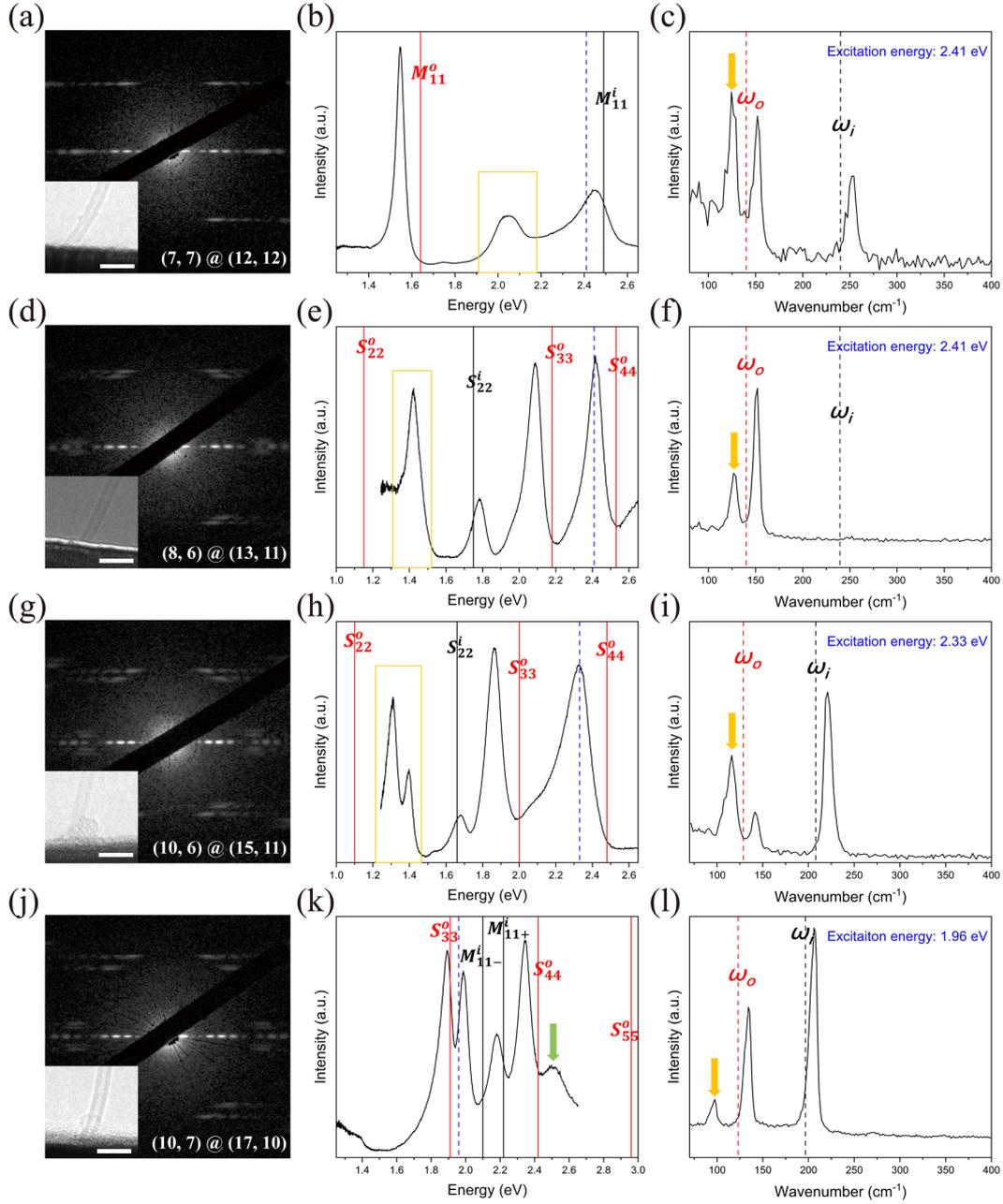

**FIG.2. Emergent phonons in commensurate and incommensurate DWNTs.** Electron diffraction pattern, Rayleigh and resonant Raman spectra for (7, 7) @ (12, 12) (a, b, c), (8, 6) @ (13, 11) (d, e, f), (10, 6) @ (15, 11) (g, h, i) and (10, 7) @ (17, 10) (j, k, l). The inset of (a,d,g,j) is the corresponding TEM image. Scale bar, 5nm. In Rayleigh spectra (b, e, h, k), the electronic transitions of pristine inner and outer nanotubes are indicated by the black and red solid lines, respectively where the superscripts, "i" and "o", represent inner and outer nanotubes, respectively. In resonant Raman spectra (c, f, i, l), the RBM frequencies of pristine inner and outer nanotubes are indicated by the black ($\omega_i$) and red ($\omega_o$) dashed lines, respectively; the excitation energies are indicated by the blue dashed lines in the corresponding Rayleigh spectra. The gold frames and arrows in Fig. 2 indicate the unusual electronic transitions and new phonon modes arising from strong inter-tube coupling. The origin of the peak indicated by the green arrow in (k) is not entirely certain (main text).

shows the resonant Raman spectrum of (7,7) @ (12,12) at low frequency region excited

at 2.41 eV (blue dashed line in Fig. 2(b)). Notably we observe three pronounced Raman peaks at ~126.7, ~151.9 and ~252.0 cm$^{-1}$, rather than two as observed in Fig. 1(b). Two peaks at higher frequency are the in-phase and out-of-phase RBM oscillations and they exhibit frequency blueshifts of ~11.8 and ~11.8 cm$^{-1}$, respectively. The strong Raman peak at ~126.7 cm$^{-1}$ (indicated by the gold arrow) cannot be explained by the coupled oscillator model since it would only yield two blue-shifted phonon modes. We also we find that this new phonon mode is persistent without showing frequency shift when excited at a different energy (Supplementary Fig. 2). The Raman G mode of (7,7) @ (12,12) is shown in Supplementary Fig. 3.

A new phonon mode at the same frequency of ~126.7 cm$^{-1}$ (marked by the gold arrow in Fig. 2(f)) is observed in an incommensurate DWNT (8,6) @ (13,11) (N = 6) that has almost identical diameter as (7,7) @ (12,12). This DWNT also shows signature of electronic "strong-coupling" in its Rayleigh spectrum as marked by the gold frame in Fig. 2(e) (Supplementary Note 3). It hints at the same physics at work for the new emergent phonon modes in Figs. 2(c) and 2(f). The coupled in-phase RBM oscillation shows a frequency blueshift of ~12.0 cm$^{-1}$, whereas the out-of-phase counterpart is not observed presumably because the excitation (blue dashed line in Fig. 2(e)) is not in resonance with that the inner (8,6). We show its Raman G mode in Supplementary Fig. 4.

We also observe new emergent phonon modes in extra two incommensurate DWNTs with larger diameters; they occur at ~116. 1 cm$^{-1}$ for (10,6) @ (15,11) (Fig. 2(i), N = 9) and at ~97.5 cm$^{-1}$ for (10,7) @ (17,10) (Fig. 2(l), N = 7). Further data including different laser excitation, Raman G-mode as well as polarization-dependent Raman measurements are shown in Supplementary Figs. 5, 6, 7. DWNT (10,6) @ (15,11) also belongs to electronic "strong-coupling" case as highlighted by the gold frame in Fig. 2(h), while the origin of the highest energy peak in Fig. 2(k) (indicated by the green arrow) is not entirely known and it can arise from strong electronic coupling if it is not a phonon side band. Each Raman spectrum shown in Fig. 2 is reproducible and solely from individual DWNT under investigation (Supplementary Fig. 8).

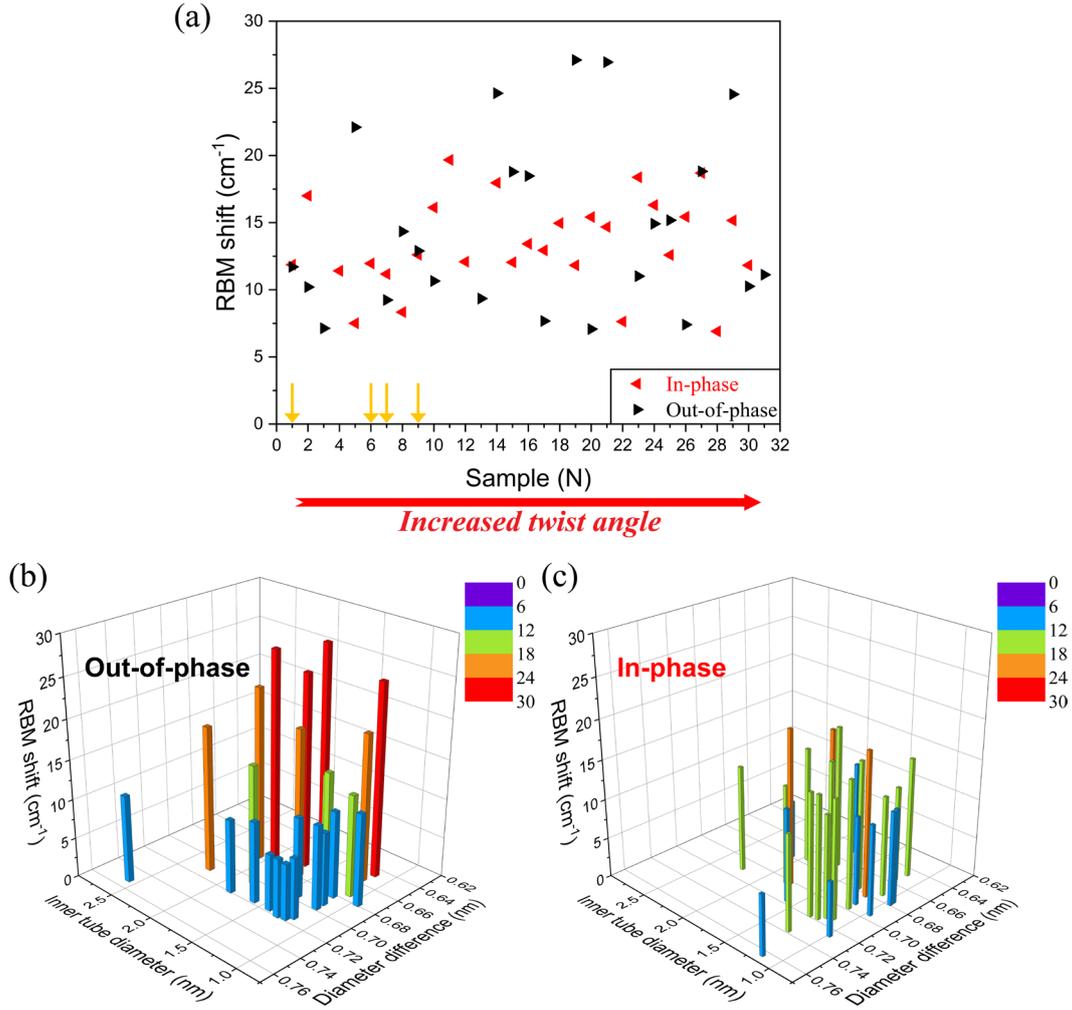

**FIG.3. Statistics of RBM oscillations for 31 DWNTs.** (a) Frequency shifts of RBM oscillations with respect to pristine SWNTs with increased twist angle from 0° (N = 1) to 23.02° (N = 31). The four DWNTs, that show new Raman peaks, are indicated by gold arrows. (b, c) Frequency shifts of out-of-phase (high frequency) and in-phase (low frequency) RBM oscillations as a function of inner tube diameter and diameter difference between inner and outer tubes.

In this study, we scrutinize 31 individual DWNTs with twist angle ranging from 0° (for N = 1) to 23.02° (for N = 31), with inner tube diameter from 0.95 to 2.58 nm, and with diameter difference from 0.647 to 0.752 nm. Twist angle is defined as the absolute chiral angle difference between two constituent inner and outer nanotubes forming a DWNT. The observed RBM shifts with respect to those in pristine SWNTs are summarized in Figs. 3(a-c). 4 out of 31 DWNTs showing the new phonons in Fig. 2 are marked by the gold arrows in Fig. 3(a). A complete dataset for 31 DWNTs are summarized in Supplementary Table 1. The blueshifts between +5 cm$^{-1}$ and +30 cm$^{-1}$ (Fig. 3(a)) and the displayed trend in Figs. 3(b) and 3(c) are consistent with previous literatures [22-27]. More detailed discussion can be found in Supplementary Note 4.

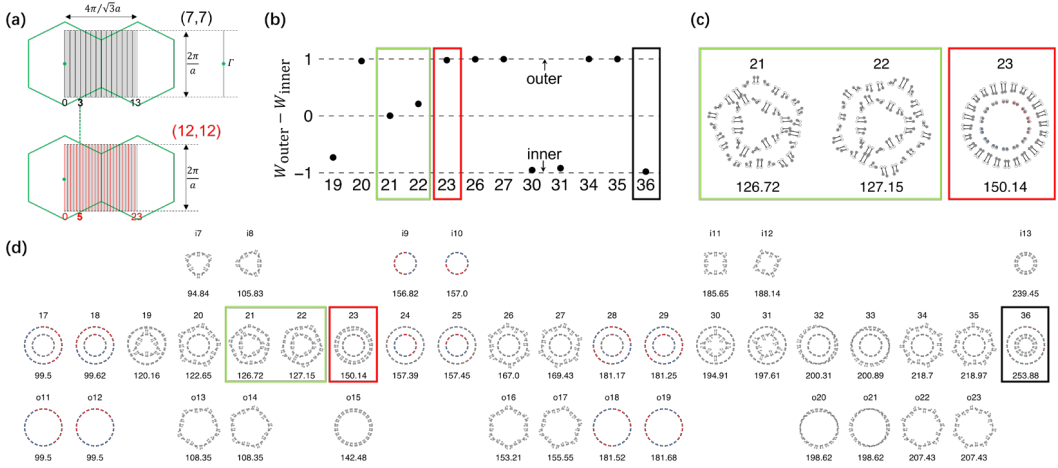

**FIG.4. DFT calculations on the phonon eigenmodes and eigenvalues for (7,7) @ (12,12).** (a) Zone folding (ZF) representation for the first Brillouin zone of (7,7) and (12,12). The length of graphene unit cell is denoted by a. (b) Calculated weight distribution ($W$), where $W \sim 0$ is found only for modes 21 and 22 (marked by the green frame), indicative of a strongly hybridized character. (c) Zoomed view for eigenmodes that are marked by the green and red frames in (d). (d) Calculated phonon eigenmodes and eigenvalues. The coupled eigenmodes are shown in the middle row, while those for pristine inner and outer nanotubes are shown in the top and bottom rows (represented by "i" and "o", respectively). The new phonon mode observed in experiment (Fig. 2(c)), in-phase (low frequency) and out-of-phase (high frequency) RBM oscillations are framed in green, red and black, respectively.

We carry out DFT calculations on the commensurate (7, 7) @ (12, 12) to obtain its phonon eigenmodes and eigenvalues between 95 cm$^{-1}$ and 255 cm$^{-1}$ as shown in Fig. 4(d). The coupled vibrational eigenmodes and eigenvalues are shown in the middle row, while those for the pristine inner and outer nanotubes are shown in the top and bottom rows; the stick lengths displayed in the eigenmodes are proportional to the maximum atomic displacements. Figure 4(c) highlights some of the eigenmodes shown in Fig. 4(d) that are directly relevant to the experiment (Fig. 2(c)). We note that our calculated eigenvalues in Fig. 4(d) quantitatively match the experimental results in Fig. 2(c) within a difference of 2 cm$^{-1}$ (Supplementary Note 5), confirming the high fidelity and accuracy for both experiment and theory.

Figure 4(d) shows that two phonons at ~126.7 (21) and ~127.2 cm$^{-1}$ (22) (also see Fig. 4(c)) closely match the experimentally observed one at ~126.7 cm$^{-1}$ (gold arrow in Fig. 2(c)). It reveals that modes 21 and 22 result primarily from a mode coupling between i8 and o13/o14 that vibrate at very close frequencies (i7 makes a small contribution). A careful mode examination shows that i7 and i8 for (7,7) constitute approximately three periods along the tube circumference, whereas o13 and o14 for (12,12) comprise approximately five periods. They are 1D quantization of the out-of-plane transverse acoustic modes (ZA modes) of 2D graphene, with $l = 3$ for inner and $l = 5$ for outer nanotubes, respectively (Fig. 4(a)), where $l$ is the angular quantum number, one of the three integers in $(q, l, j)$ specifying a phonon state in a carbon nanotube if assuming the zone folding (ZF) scheme (End Matter) [41-43].

| chirality | inner $d$ (nm) | outer $d$ (nm) | twist angle (°) | diameter ratio | $l \times \frac{2}{d}$ in/out (nm$^{-1}$) | new mode (cm$^{-1}$) |
|---|---|---|---|---|---|---|
| (7, 7) @ (12, 12) | 0.949 | 1.628 | 0 | 0.583 | 6.322/6.143 | 126.7 |
| (8, 6) @ (13, 11) | 0.953 | 1.629 | 1.96 | 0.585 | 6.296/6.139 | 126.7 |
| (10, 6) @ (15, 11) | 1.096 | 1.770 | 3.13 | 0.619 | 5.474/5.650 | 116.1 |
| (10, 7) @ (17, 10) | 1.159 | 1.851 | 2.69 | 0.626 | 5.177/5.402 | 97.5 |

**Table 1.** Summary of four DWNTs showing new phonon modes.

Table 1 summarizes the four DWNTs showing new phonon modes in this work: (i) all diameter ratios are close to $\frac{3}{5}$; (ii) the $l \times \frac{2}{d}$ values (the vertical distance from $\Gamma$ point to the $l$-th cutting line (Fig. 4(a))) are very close between inner and outer nanotubes in each DWNT and they decrease with increasing diameters; (iii) new phonon modes decrease in frequencies with increasing DWNT diameters and with decreasing averaged $l \times \frac{2}{d}$ values; (iv) all the four DWNTs have relatively small twist angles, guaranteeing parallel cutting lines of phonon states between constituent inner and outer nanotubes. All these evidences suggest a strong phonon-phonon coupling picture: a phonon state of inner SWNT at $(0, 3, j)$ strongly hybridizes with another phonon state of outer SWNT at $(0, 5, j)$. Supplementary Figure 9 shows the $l$-dependence of the calculated eigenvalues for ZA modes in pristine (7,7) and (12,12).

This coupling is very strong because the two interacting phonon states are nearly degenerate in both energy and momentum. As a measure of the coupling strength, Fig. 4(b) plots the phonon eigenmode weight distribution $W$ for selected modes (values are shown in Supplementary Table 2). Writing the displacement of the $i$-th atom (relative to its equilibrium position) as $\vec{u}_i$, we define $W = W_\text{outer} - W_\text{inner}$, with $W_\text{outer} = \sum_{i \in \text{outer}} |\vec{u}_i|^2$ and $W_\text{inner} = \sum_{i \in \text{inner}} |\vec{u}_i|^2$. Assuming phonon eigenmode is normalized as $\sum_i |\vec{u}_i|^2 = 1$, and then $W \sim 0$ indicates strong state mixing between inner and outer nanotubes, whereas $W \sim +1$ corresponds to pure outer mode and that $W \sim -1$ corresponds to pure inner mode. The new modes 21 and 22 yield $W \sim 0.002$ and $0.210$, respectively, and that they exhibit evenly distributed vibrational amplitudes over inner and outer SWNTs (Fig. 4(c)), indicative of highly hybridized nature. By contrast, the conventional in-phase (23) and out-of-phase (36) RBM oscillations show $W \sim 0.980$ and $-0.981$, respectively, indicating motions predominantly from a single wall and a weakly-coupled nature.

We highlight the experimental observation of strong Raman signals (that are comparable to conventional RBMs) for the new phonon modes. The presence of new phonon modes intimately relates to the two nearly-degenerate parent phonons (in this case, ZA phonons) in constituent inner and outer nanotubes in which their energies are diameter-dependent and they are usually not Raman-active. Our experiment indicates that the strong phonon hybridization confined in 1D can lead to Raman-active hybridized modes (Supplementary Table 2), which is not present in 2D multilayer graphene [44]. We further theoretically investigate the phonon eigenmodes and eigenvalues of extra two commensurate DWNTs (10,10) @ (15,15) and (5,5) @ (10,10) with a diameter ratio of 2/3 and 1/2, respectively (Supplementary Fig. 10 and Fig. 11), whose results further support the proposed resonant coupling mechanism and infer an additional curvature effect beyond quantum confinement we have considered so far (End Matter). Further discussions and insights on the strongly hybridized 1D phonons are included in End Matter.

In conclusion, we discover strongly hybridized 1D phonons in both commensurate and incommensurate DWNTs. A nearly-degenerate phonon coupling picture is proposed to explain the experiments. Our work unlocks exciting new opportunities to explore and engineer phonon hybridizations in 1D systems.

**Acknowledgements：**

We acknowledge Saito Riichiro for insightful discussions. This work is mainly supported by National Key R&D Program of China (2022YFA1203400), National Natural Science Foundation of China (12174335), Zhejiang Provincial Natural Science Foundation of China (LR23A040002) and National Key R&D Program of China (2023YFA1407900). Parts of this study are supported by JSPS KAKENHI (JP24K06968 to T.K, 23K23161 to D.K, JP24H02218 and JP23H05469 to R.K.). P.M. acknowledges the financial support from the National Natural Science Foundation of China (12074260) and the NYU-ECNU Institute of Physics at NYU Shanghai. The calculations in this study have been done using the Numerical Materials Simulator at NIMS, and using the facilities of the Supercomputer Center, the Institute for Solid State Physics, the University of Tokyo.

**Reference:**
[1] C. Kittel, *Introduction to Solid State Physics* (Wiley, New York, 2005).
[2] J. Bardeen, L. N. Cooper, and J. R. Schrieffer, Theory of Superconductivity, Physical Review **108**, 1175 (1957).
[3] S. Dai, Z. Fei, Q. Ma, A. S. Rodin, M. Wagner, A. S. McLeod, M. K. Liu, W. Gannett, W. Regan, K. Watanabe *et al.*, Tunable Phonon Polaritons in Atomically Thin van der Waals Crystals of Boron Nitride, Science **343**, 1125 (2014).
[4] R. Peierls, Zur kinetischen Theorie der Wärmeleitung in Kristallen, Annalen der Physik **395**, 1055 (1929).
[5] X. Qian, J. Zhou, and G. Chen, Phonon-engineered extreme thermal conductivity materials, Nature Materials **20**, 1188 (2021).
[6] O. Delaire, J. Ma, K. Marty, A. F. May, M. A. McGuire, M. H. Du, D. J. Singh, A. Podlesnyak, G. Ehlers, M. D. Lumsden *et al.*, Giant anharmonic phonon scattering in PbTe, Nature Materials **10**, 614 (2011).
[7] B. Poudel, Q. Hao, Y. Ma, Y. Lan, A. Minnich, B. Yu, X. Yan, D. Wang, A. Muto, D. Vashaee *et al.*, High-Thermoelectric Performance of Nanostructured Bismuth Antimony Telluride Bulk Alloys, Science **320**, 634 (2008).
[8] O. Hellman and D. A. Broido, Phonon thermal transport in $Bi_2Te_3$ from first principles, Physical Review B **90**, 134309 (2014).
[9] M.-L. Lin, Q.-H. Tan, J.-B. Wu, X.-S. Chen, J.-H. Wang, Y.-H. Pan, X. Zhang, X. Cong, J. Zhang, W. Ji *et al.*, Moiré Phonons in Twisted Bilayer $MoS_2$, ACS Nano **12**, 8770 (2018).
[10] M. Koshino and Y.-W. Son, Moiré phonons in twisted bilayer graphene, Physical Review B **100**, 075416 (2019).

[11] J. Quan, L. Linhart, M.-L. Lin, D. Lee, J. Zhu, C.-Y. Wang, W.-T. Hsu, J. Choi, J. Embley, C. Young *et al.*, Phonon renormalization in reconstructed $MoS_2$ moiré superlattices, Nature Materials **20**, 1100 (2021).

[12] L. P. A. Krisna and M. Koshino, Moiré phonons in graphene/hexagonal boron nitride moiré superlattice, Physical Review B **107**, 115301 (2023).

[13] J. M. Luttinger, An Exactly Soluble Model of a Many-Fermion System, Journal of Mathematical Physics **4**, 1154 (1963).

[14] M. Bockrath, D. H. Cobden, J. Lu, A. G. Rinzler, R. E. Smalley, L. Balents, and P. L. McEuen, Luttinger-liquid behaviour in carbon nanotubes, Nature **397**, 598 (1999).

[15] R. E. Peierls, *Quantum Theory of Solids* (Clarendon Press, Oxford, 1996).

[16] A. Luther and I. Peschel, Single-particle states, Kohn anomaly, and pairing fluctuations in one dimension, Physical Review B **9**, 2911 (1974).

[17] K. Liu, C. Jin, X. Hong, J. Kim, A. Zettl, E. Wang, and F. Wang, Van der Waals-coupled electronic states in incommensurate double-walled carbon nanotubes, Nature Physics **10**, 737 (2014).

[18] S. Zhao, T. Kitagawa, Y. Miyauchi, K. Matsuda, H. Shinohara, and R. Kitaura, Rayleigh scattering studies on inter-layer interactions in structure-defined individual double-wall carbon nanotubes, Nano Research **7**, 1548 (2014).

[19] G. Gordeev, S. Wasserroth, H. Li, A. Jorio, B. S. Flavel, and S. Reich, Dielectric Screening inside Carbon Nanotubes, Nano Letters **24**, 8030 (2024).

[20] M. Koshino, P. Moon, and Y.-W. Son, Incommensurate double-walled carbon nanotubes as one-dimensional moiré crystals, Physical Review B **91**, 035405 (2015).

[21] S. Zhao, P. Moon, Y. Miyauchi, T. Nishihara, K. Matsuda, M. Koshino, and R. Kitaura, Observation of Drastic Electronic-Structure Change in a One-Dimensional Moiré Superlattice, Physical Review Letters **124**, 106101 (2020).

[22] D. Levshov, T. X. Than, R. Arenal, V. N. Popov, R. Parret, M. Paillet, V. Jourdain, A. A. Zahab, T. Michel, Y. I. Yuzyuk *et al.*, Experimental Evidence of a Mechanical Coupling between Layers in an Individual Double-Walled Carbon Nanotube, Nano Letters **11**, 4800 (2011).

[23] K. Liu, X. Hong, M. Wu, F. Xiao, W. Wang, X. Bai, J. W. Ager, S. Aloni, A. Zettl, E. Wang *et al.*, Quantum-coupled radial-breathing oscillations in double-walled carbon nanotubes, Nature Communications **4**, 1375 (2013).

[24] G. Gordeev, S. Wasserroth, H. Li, B. Flavel, and S. Reich, Moiré-Induced Vibrational Coupling in Double-Walled Carbon Nanotubes, Nano Letters **21**, 6732 (2021).

[25] V. N. Popov and L. Henrard, Breathinglike phonon modes of multiwalled carbon nanotubes, Physical Review B **65**, 235415 (2002).

[26] E. Dobardžić, J. Maultzsch, I. Milošević, C. Thomsen, and M. Damnjanović, The radial breathing mode frequency in double-walled carbon nanotubes: an analytical approximation, physica status solidi (b) **237**, R7 (2003).

[27] A. Rahmani, J. L. Sauvajol, J. Cambedouzou, and C. Benoit, Raman-active modes in finite and infinite double-walled carbon nanotubes, Physical Review B **71**, 125402 (2005).

[28] See Supplementary Note 1 in Supplemental Material for detailed discription of our experimental methods and DFT calculations, which includes Refs. [29-34].

[29] Ceresoli, G. L. Chiarotti, M. Cococcioni, I. Dabo et al., QUANTUM ESPRESSO: a modular and open-source software project for quantum simulations of materials, Journal of Physics: Condensed Matter **21**, 395502 (2009).

[30] P. Giannozzi, O. Andreussi, T. Brumme, O. Bunau, M. Buongiorno Nardelli, M. Calandra, R. Car, C. Cavazzoni, D. Ceresoli, M. Cococcioni et al., Advanced capabilities for materials modelling with Quantum ESPRESSO, Journal of Physics: Condensed Matter **29**, 465901 (2017).

[31] A. Togo, L. Chaput, T. Tadano, and I. Tanaka, Implementation strategies in phonopy and phono3py, Journal of Physics: Condensed Matter **35**, 353001 (2023).

[32] A. Togo, First-principles Phonon Calculations with Phonopy and Phono3py, Journal of the Physical Society of Japan **92**, 012001 (2022).

[33] A. Dal Corso, Pseudopotentials periodic table: From H to Pu, Computational Materials Science **95**, 337 (2014).

[34] S. Grimme, J. Antony, S. Ehrlich, and H. Krieg, A consistent and accurate ab initio parametrization of density functional dispersion correction (DFT-D) for the 94 elements H-Pu, The Journal of Chemical Physics **132**, 154104 (2010).

[35] M. Kociak, K. Suenaga, K. Hirahara, Y. Saito, T. Nakahira, and S. Iijima, Linking Chiral Indices and Transport Properties of Double-Walled Carbon Nanotubes, Physical Review Letters **89**, 155501 (2002).

[36] K. Liu, Z. Xu, W. Wang, P. Gao, W. Fu, X. Bai, and E. Wang, Direct determination of atomic structure of large-indexed carbon nanotubes by electron diffraction: application to double-walled nanotubes, Journal of Physics D: Applied Physics **42**, 125412 (2009).

[37] K. Liu, J. Deslippe, F. Xiao, R. B. Capaz, X. Hong, S. Aloni, A. Zettl, W. Wang, X. Bai, S. G. Louie *et al.*, An atlas of carbon nanotube optical transitions, Nature Nanotechnology **7**, 325 (2012).

[38] K. Liu, W. Wang, M. Wu, F. Xiao, X. Hong, S. Aloni, X. Bai, E. Wang, and F. Wang, Intrinsic radial breathing oscillation in suspended single-walled carbon nanotubes, Physical Review B **83**, 113404 (2011).

[39] See Supplementary Fig. 1 for simulated result, which includes Ref. [40].

[40] M. Gao, J. M. Zuo, R. Zhang, and L. A. Nagahara, Structure determinations of double-wall carbon nanotubes grown by catalytic chemical vapor deposition, Journal of Materials Science **41**, 4382 (2006).

[41] S. Piscanec, M. Lazzeri, J. Robertson, A. C. Ferrari, and F. Mauri, Optical phonons in carbon nanotubes: Kohn anomalies, Peierls distortions, and dynamic effects, Physical Review B **75**, 035427 (2007).

[42] R. Saito, G. Dresselhaus, and M. S. Dresselhaus, *Physical Properties of Carbon Nanotubes* (Imperial College Press, London, 1998).

[43] M. S. Dresselhaus and P. C. Eklund, Phonons in carbon nanotubes, Advances in Physics **49**, 705 (2000).

[44] L. M. Malard, D. C. Elias, E. S. Alves, and M. A. Pimenta, Observation of Distinct Electron-Phonon Couplings in Gated Bilayer Graphene, Physical Review Letters **101**, 257401 (2008).

**End Matter:**

*Specify a phonon state in a carbon nanotube within zone folding scheme*

Translational and rotational boundary conditions allow a phonon of an isolated SWNT to be specified by three quantum numbers $(q, l, j)$. Here $q$ is the wave number associated with the translational periodicity along the tube axis, $l$ is the angular quantum number associated with the rotational periodicity along the circumferential direction and it takes on $N$ integer values $l = 0, 1, 2, \ldots N-1$ with $N$ being the number of hexagons in the SWNT unit cell, $j$ is one of the six phonon branches for a given $(q, l)$ in 2D graphene Brillouin zone [41]. For an armchair nanotube with the chirality of $(n,n)$, $N = 2n$.

*Curvature effect beyond quantum confinement*

Although all the new phonon modes reported in this work are founded to have a diameter ratio close to $\frac{3}{5}$, the rich body of DWNT crystals may allow an abundance of strong phonon-phonon coupling with diameter ratios other than $\frac{3}{5}$. We theoretically investigate the phonon eigenmodes and eigenvalues of another hypothesized yet realistic commensurate DWNT (5,5) @ (10,10) with a diameter ratio of $\frac{1}{2}$ (Supplementary Fig. 11). The strong phonon-phonon coupling is not present in this case because of the large energy mismatch. This hints that the curvature effect plays a role, especially for small-diameter nanotubes, beyond the quantum confinement we have considered so far (also see Supplementary Fig. 9).

*Symmetry argument for the new phonon mode*

The phonon mode shapes (21 and 22 in Figs. 4(c) and 4(d)) infer that the strongly hybridized modes are Raman active symmetrywise. Firstly, we notice that the in-phase RBM (23) gets distorted in a 5-fold (pentagonal) rotational symmetric way, which indicates the symmetry of DWNT allows mixing between pure RBM (inner nanotube, in this case) and pentagonal modes. (The pentagonal distortion was found in a previous work without explanation [25]. A possible explanation for the pentagonal distortion is described in Supplementary Note 6.) On the other hand, we have seen that the strongly hybridized modes show pentagonal shapes, which infers that these modes can also be Raman active. Note that 7 and 12 have no common divisor, and therefore, there is no rotational symmetry (even discrete ones) about the axis of (7,7) @ (12,12) in a strict sense. This allows modes with different angular momenta to couple in principles, but the coupling effect is prominent only when $l$ satisfies a matching condition as explained above. Inversely speaking, if nanotubes are modeled as continuum elastic tubes, pure RBMs cannot mix with pentagonal modes, since they are distinguished by the continuum rotational symmetry. This could suggest importance of microscopic electronic structures and a complex interplay between the phononic and electronic structures, in accordance with our observation that new Raman mode is attendant with strong coupling features in Rayleigh spectrum.

*Other intriguing questions for future studies*

The exact connection and interplay between the strong electronic coupling and strong phononic coupling are not fully understood yet. One point we can draw at this moment is that all DWNTs satisfying the "strong-coupling" condition show signs of the strong phonon-phonon coupling in this work.

The synergistic quantum confinement and curvature effects were proposed to make the atomic displacements aligned along the tube axis (LO phonons) and tube circumference (TO phonons) in chiral nanotubes [41]. This intriguing preference may facilitate the strong phonon coupling effect in incommensurate DWNTs with small twists. Understanding the precise role of moiré effect and being capable of determining the nanotube handedness in future work will enhance the in-depth understanding of this topic.

# Supplemental Material
# Strongly hybridized phonons in one-dimensional van der Waals crystals

**Supplementary Note 1.**

To investigate the effect of strong phonon coupling in DWNT crystals, we directly grew air-suspended DWNTs with high quality across an open slit (~30 μm in width) by alcohol catalytic chemical vapor deposition (ACCVD). Combined Rayleigh scattering and resonant Raman scattering spectroscopies measured in the same configuration were employed to probe the electronic optical transitions and phonon vibrations for chirality-resolved individual DWNTs. We focused either a supercontinuum laser source (1.2–2.65 eV for Rayleigh) or monochromic laser lines (514, 532, 561 and 633 nm for Raman) on the central part of each suspended DWNT, with the light polarized along each nanotube axis. The light scattered by the nanotube was collected and directed to a CCD camera and a spectrometer. The Rayleigh spectra normalized by the incident light were further corrected by multiplying a factor of $1/E^3$ for different photon energy $E$ to directly reflect the susceptibility of nanotube dipoles. The spectral resolutions for Rayleigh and Raman measurements were about 5 meV and 1–2 cm$^{-1}$, respectively. The structure (i.e., the chirality) of each individual DWNT was determined by nanobeam electron diffraction in a transmission electron microscope (TEM) operated at 80 keV (JEM-F200) where diffractions at two ends of the same DWNT were measured to ensure the structural homogeneity and unambiguous assignment over the whole suspended part, further consolidated by simulations. Since electron diffraction cannot distinguish the left-hand species from right-hand ones of chiral SWNTs, we assumed that each DWNT in this study contain two constituent SWNTs with same handedness.

The phonon modes are derived using the density functional theory through Quantum Espresso [1,2] and Phonopy [3,4] packages. First, relaxed lattice structures of single- and double-walled carbon nanotubes have been obtained with the criterion that the maximum force on the atoms went below 0.001 Ry/bohr. In our calculation, we have adapted the projector augmented wave method, taken a pseudopotential from [3,4] and used the PBE-GGA functional [5]. The van der Waals forces were included through DFT-D3 method [6]. The cutoff energies for the wave functions and the charge density were set to 60 Ry and 489 Ry, respectively. The nanotube axis was aligned with z-axis, and for x- and y- directions, the system was assumed to be periodic with the lattice constant 4 nm (and 4.5 nm only for the case with the (10,10) @ (15,15) case), which was sufficiently larger than the nanotube diameter. Then, the force constants required for the phonon mode calculation was obtained by perturbing the atomic positions in $1 \times 1 \times 4$ supercell of the above-mentioned relaxed structures for single- and double-walled carbon nanotubes. (The perturbed structures were automatically generated by Phonopy.) The phonon frequencies and mode shapes were given by diagonalizing dynamical matrices consists of the force constants.

**Supplementary Note 2.**

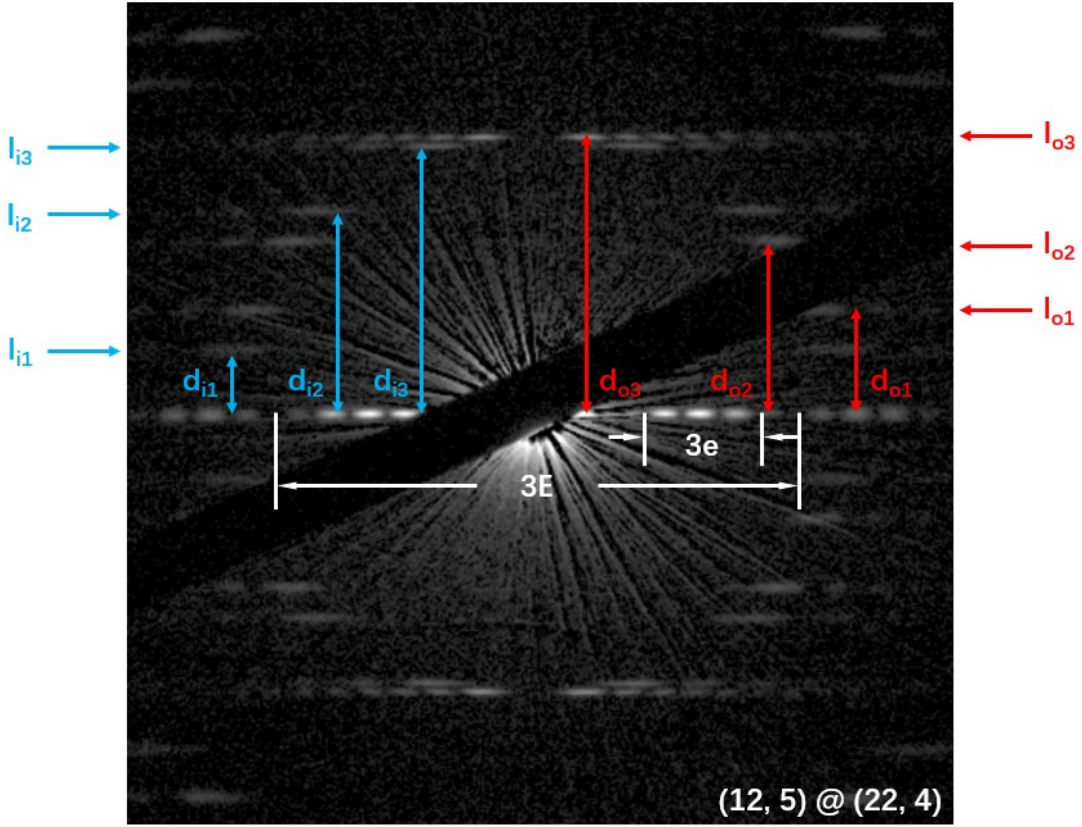

The electron diffraction pattern of a DWNT contains its structural information, from which the chiral indices $(n_i, m_i)$ @ $(n_o, m_o)$ can be determined unambiguously. Taking (12, 5) @ (22, 4) in the main text (Fig. 1(a)) as an example, the intensity distribution of the equatorial line oscillates with a small period $e \propto 1/\overline{D}$, within an oscillatory envelope of a large period $E \propto 1/\delta D$, where $\overline{D}$ and $\delta D$ are the average diameter and diameter difference of inner and outer nanotubes, respectively. Moreover, there are two sets of layer lines, [$l_{i1}$, $l_{i2}$, $l_{i3}$] and [$l_{o1}$, $l_{o2}$, $l_{o3}$], in the electron diffraction pattern, originating from inner and outer nanotubes, respectively. The distances between these layer lines and the equatorial line ($d_{i1}$, $d_{i2}$, $d_{i3}$) for inner SWNT and ($d_{o1}$, $d_{o2}$, $d_{o3}$) for outer SWNT, along with the periods of $e$ and $E$, are dicisive experimental inputs obtained from the analysis of electron diffraction to determine the chiral indices $(n_i, m_i)$ @ $(n_o, m_o)$ [7]:

$$\frac{n_o}{\cos\tau} = \frac{\pi}{\sqrt{3}}(2d_{o3} - d_{o2})\left(\frac{1}{e} + \frac{1}{E}\right) \tag{1}$$

$$\frac{m_o}{\cos\tau} = \frac{\pi}{\sqrt{3}}(2d_{o2} - d_{o3})\left(\frac{1}{e} + \frac{1}{E}\right) \tag{2}$$

$$\frac{n_i}{\cos\tau} = \frac{\pi}{\sqrt{3}}(2d_{i3} - d_{i2})\left(\frac{1}{e} - \frac{1}{E}\right) \tag{3}$$

$$\frac{m_i}{\cos\tau} = \frac{\pi}{\sqrt{3}}(2d_{i2} - d_{i3})\left(\frac{1}{e} - \frac{1}{E}\right) \tag{4}$$

where τ is the tilt angle correction of the nanotube at the substrate edge where the diffraction pattern is taken, and it is usually less than 30°. Using the euqations (1)-(4), we can get the chirality candidates of the DWNT being investigated first. After comparing the experimental result with the simulated diffraction patterns of these candidates, we could rule out other possibilities and uniquely determine the chiral indices of the DWNT to be (12, 5) @ (22, 4). The simulated diffration patterns of two candidates are shown below, in which (12, 5) @ (22, 4) agrees well the the experiment.

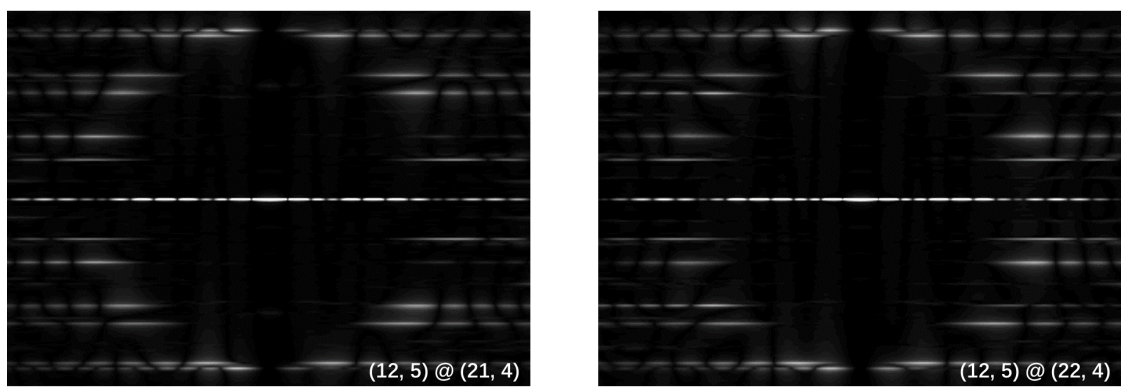

**Supplementary Figure 1.**

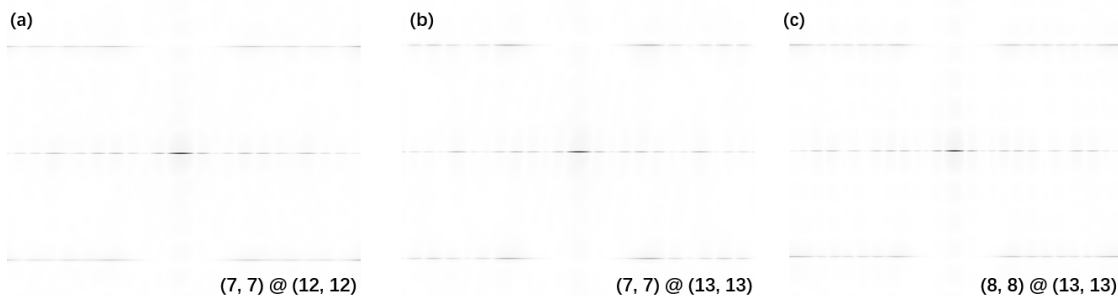

**Fig. S1. Simulated diffraction patterns of candidate DWNTs by using an analytical approach [8].** (a) (7, 7) @ (12, 12), (b) (7, 7) @ (13, 13), (c) (8, 8) @ (13, 13). It is apparent that only (a) matches the experimental result shown in Fig. 2(a) in the main text.

**Supplementary Figure 2.**

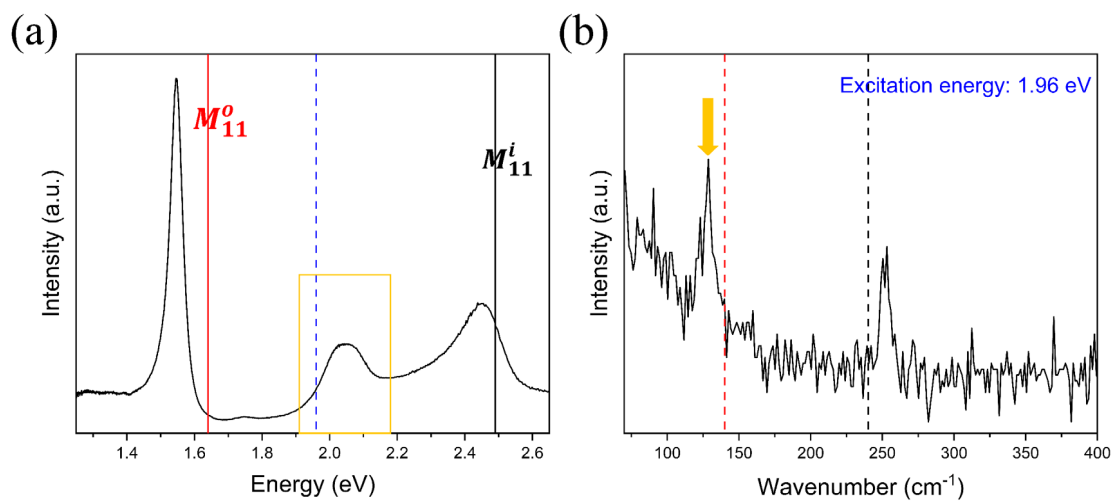

**Fig. S2. The Raman spectrum (b) of (7, 7) @ (12, 12) in the low frequency region with the excitation energy of ~1.96 eV indicated by the blue dashed line in its Rayleigh spectrum (a).** The excitation energy is close to the emerging optical transitions that cannot be attributed to electronic transitions of neither inner nor outer nanotubes. The new phonon mode, indicated by gold arrow in Fig. S2(b), is still present without frequency shift, though the in-phase RBM oscillation is absent, primarily owing to the weak intensity. The RBM oscillations for pristine inner and outer nanotubes are indicated by the black and red dashed lines, respectively.

**Supplementary Figure 3.**

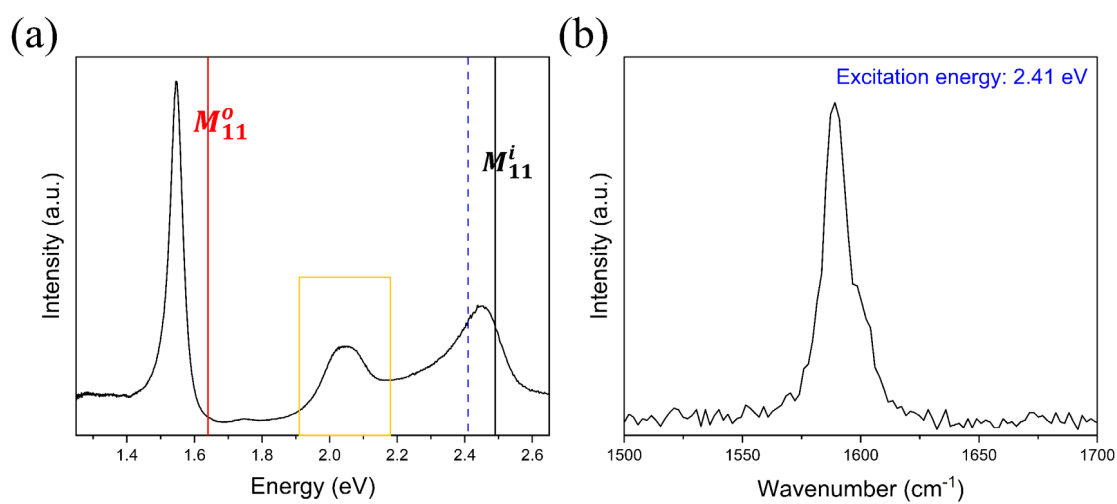

**Fig. S3. The resonant Raman spectrum (b) of (7, 7) @ (12, 12) in the G-band region with the excitation energy of ~2.41 eV (same as in Fig. 2(c) in the main text), indicated by the blue dashed line in its Rayleigh spectrum (a).** The line shape of G mode is asymmetric, and contains multiple peaks.

## Supplementary Note 3.

The "strong-coupling" condition for electronic inter-tube interactions in DWNTs is that the chiral vectors of inner and outer nanotubes, $C_{in}$ and $C_{out}$, are nearly parallel to each other and, at the same time, the difference of the two vectors, $C_{out}$ - $C_{in}$, is strictly along the armchair direction [9]. As a certain DWNT matches the "strong-coupling" criterion, the electronic states of the components strongly hybridize with each other due to the presence of moiré superlattice potential, leading to strong electronic band reconstruction. For (8, 6) @ (13, 11) and (10, 6) @ (15, 11), the differences in chiral angle of the inner and outer nanotubes are ~1.96° and ~3.13°, respectively, and the chiral vector differences, $C_{out}$ - $C_{in}$, are both (5,5), parallel to the armchair direction. Therefore, (8, 6) @ (13, 11) and (10, 6) @ (15, 11) are both strongly-coupled incommensurate DWNTs, whose electronic band structures can be significantly modified by the moiré potential. Indeed, their Rayleigh spectra (Figs. 2(e) and 2(h) in the main text) exhibit pronounced deviation from the perturbation theory, manifesting as pronounced energy shift outside the range between -200 meV and +50 meV and/or the emergence of new optical transitions. For (8, 6) @ (13, 11), its Rayleigh spectrum (FIG.2 (e) in the main text) shows four optical transitions in the range of 1.25-2.65 eV. Three transitions at high energy can be assigned to those of constituent inner and outer nanotubes with energy shifts in the range between -200 meV and +50 meV. However, as the peak at ~1.42 eV (framed in gold) is assigned to $S_{22}^o$ transition of outer tube (13, 11), the energy shift is ~+270 meV, beyond the range of perturbation theory. The Rayleigh spectrum of (10, 6) @ (15, 11) shows five distinct optical resonances at about 1.31, 1.40, 1.68, 1.86 and 2.33 eV (FIG.2 (h) in the main text). The three high-energy peaks at 1.68, 1.86 and 2.33eV can be assigned to $S_{22}^i$ transition of inner tube (10, 6), $S_{33}^o$ and $S_{44}^o$ transitions of outer tube (15,11), respectively, with moderate energy shifts of about 20, -140 and -150 meV. But for the two peaks at lower energy side (framed in gold), they cannot be explained in the framework of perturbation theory. Even though the resonance at ~1.31 eV can be assigned to the $S_{22}^o$ transition of outer tube (a blueshift of about +210 meV is already beyond the perturbation theory [10]), the resonance peak at ~1.40 eV cannot be assigned to neither pristine inner nor outer nanotube. The commensurate DWNT (7,7) @ (12,12) also satisfies the "strong-coupling" condition.

## Supplementary Figure 4.

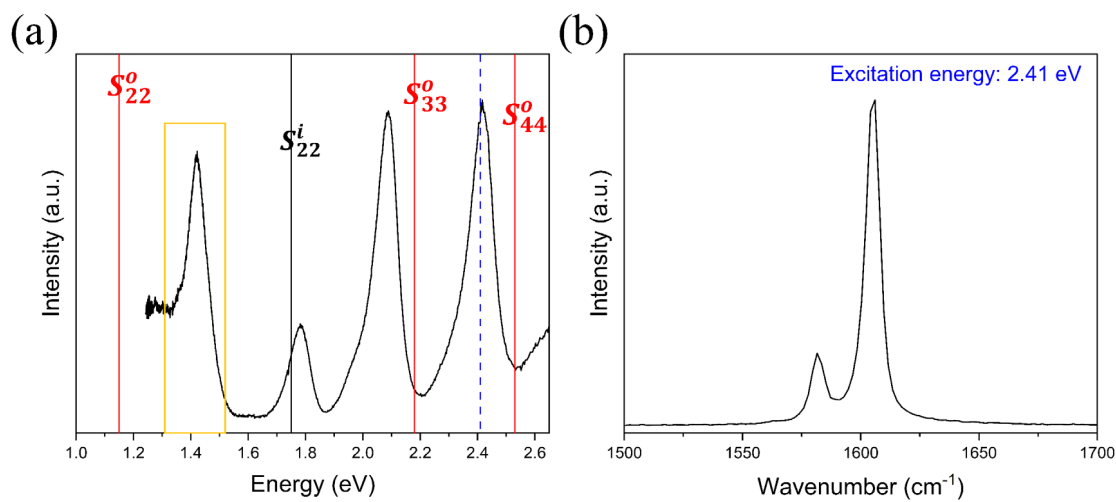

**Fig. S4. The resonant Raman spectrum (b) of (8, 6) @ (13, 11) in the G-band region.** A clear splitting of G-band is observed. The excitation energy is the same as that in Fig. 2(f) in the main text and is indicated by the blue dashed line in its Rayleigh spectrum (a).

## Supplementary Figure 5.

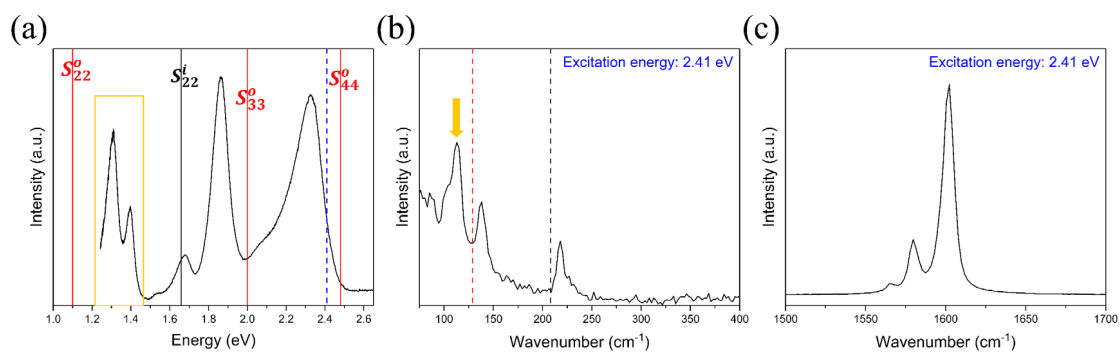

**Fig. S5. Raman measurement of DWNT (10, 6) @ (15, 11) with a different excitation energy.** The Raman spectra of (10, 6) @ (15, 11) in the low frequency (b) and G-band region (c). The excitation energy at ~514 nm is different from Fig. 2(i) in the main text and is indicated by the blue dashed line in the Rayleigh spectrum (a). The Raman spectrum in Fig. S5(b) is similar to Fig. 2(i) in the main text.

**Supplementary Figure 6.**

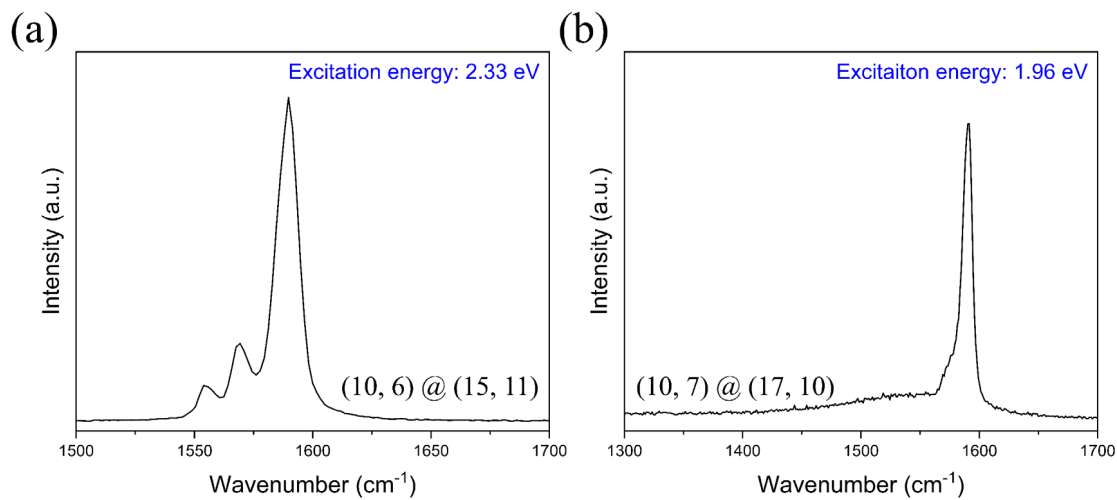

**Fig. S6. The Raman spectra of (10, 6) @ (15, 11) (a) and (10, 7) @ (17, 10) (b) in the G-band region.** The excitation energies of both DWNTs are the same as those in FIG.2 in the main text. For (10, 6) @ (15, 11), a clear splitting of G-band into multiple peaks is observed.

**Supplementary Figure 7.**

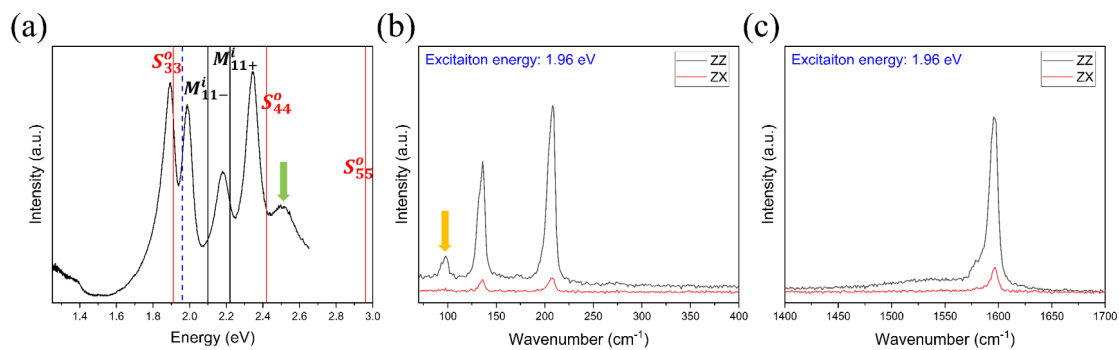

**Fig. S7. Polarization-dependent Raman measurement on DWNT (10, 7) @ (17, 10).** The Raman spectra of (10, 7) @ (17, 10) in the low frequency (b) and G-band region (c) with different polarization configurations. The excitation energy of ~1.96eV is indicated by the bule dashed line in the corresponding Rayleigh spectrum (a). The origin of the highest energy peak in (a) (indicated by the green arrow) is not entirely known and it can arise from strong electronic coupling if it is not a phonon side band. The new phonon mode in (b) shows a favorable polarization in the ZZ configuration and vanishes in the ZX configuration, where ZZ means the incident and outgoing fields are polarized along tube axis and that ZX means that the incident and outgoing fields are polarized along the tube and circumferential direction. The result means that the new phonon mode is not $E_{1g}$ symmetrywise, but likely to be $A_{1g}$ symmetrywise as for the pristine RBMs. The residue ZX signals for the in-phase and out-of-phase

oscillations (b) and G mode (c) are due to the imperfect nanotube placement with respect to the two orthogonal polarizers.

**Supplementary Figure 8.**

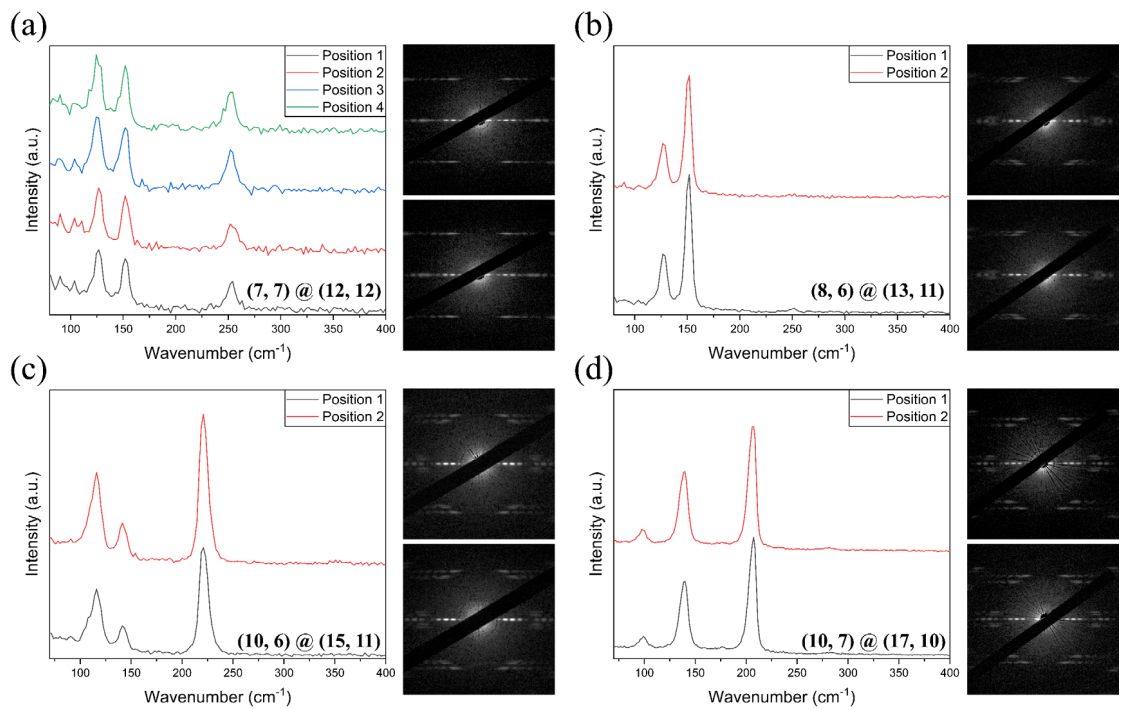

**Fig. S8. Raman spectra taken at different positions and diffraction patterns taken at two ends of the suspended nanotubes for four DWNTs exhibiting new Raman peaks.** (a), (b), (c) and (d) correspond to (7, 7) @ (12, 12), (8, 6) @ (13, 11), (10, 6) @ (15, 11) and (10, 7) @ (17, 10), respectively.

**Supplementary Table 1.**

| sample | chirality | inner tube diameter (nm) | outer tube diameter (nm) | diameter difference (nm) | chiral angle difference (°) | resonance | in-phase shift (cm$^{-1}$) | out-of-phase shift (cm$^{-1}$) | diameter ratio | new peak (cm$^{-1}$) |
|---|---|---|---|---|---|---|---|---|---|---|
| 1 | (7, 7) @ (12, 12) | 0.949 | 1.628 | 0.679 | 0 | inner tube | 11.87 | 11.71 | 0.583 | 126.7 |
| 2 | (14, 6) @ (21, 9) | 1.392 | 2.088 | 0.696 | 0 | | 17.00 | 10.21 | 0.667 | |
| 3 | (15, 5) @ (23, 7) | 1.412 | 2.129 | 0.717 | 1.02 | | -- | 7.13 | 0.663 | |
| 4 | (9, 5) @ (16, 8) | 0.962 | 1.657 | 0.695 | 1.52 | outer tube | 11.41 | -- | 0.580 | |
| 5 | (21, 9) @ (27, 13) | 2.088 | 2.767 | 0.679 | 1.58 | | 7.50 | 22.10 | 0.755 | |
| 6 | (8, 6) @ (13, 11) | 0.953 | 1.629 | 0.676 | 1.96 | outer tube | 11.96 | -- | 0.585 | 126.7 |
| 7 | (10, 7) @ (17, 10) | 1.159 | 1.851 | 0.692 | 2.69 | inner tube | 11.17 | 9.24 | 0.626 | 97.5 |
| 8 | (14, 5) @ (20, 9) | 1.336 | 2.013 | 0.677 | 2.95 | | 8.34 | 14.34 | 0.664 | |
| 9 | (10, 6) @ (15, 11) | 1.096 | 1.770 | 0.674 | 3.13 | outer tube | 12.61 | 12.89 | 0.619 | 116.1 |
| 10 | (13, 3) @ (22, 3) | 1.153 | 1.851 | 0.698 | 3.85 | outer tube | 16.12 | 10.65 | 0.623 | |
| 11 | (15, 11) @ (23, 13) | 1.770 | 2.472 | 0.702 | 4.03 | | 19.67 | -- | 0.716 | |
| 12 | (13, 5) @ (19, 10) | 1.260 | 1.998 | 0.738 | 4.23 | outer tube | 12.08 | -- | 0.631 | |
| 13 | (16, 11) @ (25, 12) | 1.841 | 2.560 | 0.719 | 5.37 | both | -- | 9.34 | 0.719 | |
| 14 | (15, 10) @ (18, 17) | 1.707 | 2.374 | 0.667 | 5.64 | | 17.96 | 24.63 | 0.719 | |
| 15 | (12, 5) @ (16, 11) | 1.185 | 1.841 | 0.656 | 7.27 | inner tube | 12.04 | 18.78 | 0.644 | |
| 16 | (17, 16) @ (27, 16) | 2.238 | 2.948 | 0.710 | 7.40 | outer tube | 13.42 | 18.47 | 0.759 | |
| 17 | (12, 5) @ (22, 4) | 1.185 | 1.899 | 0.714 | 8.42 | outer tube | 12.94 | 7.67 | 0.624 | |
| 18 | (12, 4) @ (16, 11) | 1.129 | 1.841 | 0.712 | 10.00 | outer tube | 14.95 | -- | 0.613 | |
| 19 | (18, 7) @ (29, 3) | 1.749 | 2.397 | 0.648 | 10.88 | inner tube | 11.82 | 27.11 | 0.730 | |
| 20 | (9, 9) @ (19, 9) | 1.221 | 1.939 | 0.718 | 11.65 | outer tube | 15.41 | 7.07 | 0.630 | |
| 21 | (20, 8) @ (32, 3) | 1.956 | 2.631 | 0.675 | 11.67 | both | 14.68 | 26.96 | 0.743 | |
| 22 | (10, 7) @ (21, 6) | 1.159 | 1.923 | 0.764 | 11.96 | outer tube | 7.64 | -- | 0.603 | |
| 23 | (14, 2) @ (18, 9) | 1.182 | 1.864 | 0.682 | 12.52 | inner tube | 18.38 | 11.01 | 0.634 | |
| 24 | (15, 3) @ (18, 11) | 1.308 | 1.986 | 0.678 | 13.12 | inner tube | 16.31 | 14.92 | 0.659 | |
| 25 | (18, 8) @ (31, 2) | 1.806 | 2.509 | 0.703 | 14.38 | outer tube | 12.59 | 15.17 | 0.720 | |
| 26 | (16, 1) @ (20, 9) | 1.294 | 2.013 | 0.719 | 14.65 | | 15.44 | 7.40 | 0.643 | |
| 27 | (13, 11) @ (26, 6) | 1.629 | 2.307 | 0.678 | 17.08 | | 18.69 | 18.82 | 0.706 | |
| 28 | (12, 1) @ (15, 10) | 0.981 | 1.707 | 0.726 | 19.45 | outer tube | 6.90 | -- | 0.575 | |
| 29 | (10, 7) @ (22, 2) | 1.159 | 1.806 | 0.647 | 19.87 | | 15.16 | 24.56 | 0.642 | |
| 30 | (18, 4) @ (17, 17) | 1.589 | 2.306 | 0.717 | 20.17 | both | 11.83 | 10.25 | 0.689 | |
| 31 | (21, 17) @ (41, 3) | 2.582 | 3.334 | 0.752 | 23.02 | inner tube | -- | 11.12 | 0.774 | |

**Table S1. Statistics for 31 individual DWNTs in this study.** Four DWNTs showing new phonon modes are marked in red. The excitation energy of Raman measurement can be resonant with inner tube or outer tube or both. The "resonance" condition for three DWNTs are left blank because the optical transitions from inner and outer

nanotubes are too close to be distinguishable in Rayleigh spectra. It is noticeable that 10 out of 31 DWNTs (N = 3, 4 ,6, 11, 12, 13, 18, 22, 28, 31) show only one RBM oscillation which is presumably caused by the off-resonance condition, as the excitation is only resonant with either inner or outer tube. For sample N = 13, although the excitation energy is resonant with both nanotubes, the in-phase oscillation is absent.

**Supplementary Note 4.**

RBM shifts for all DWNTs fall into a range between +5 cm$^{-1}$ and +30 cm$^{-1}$ (Fig. 3(a)). We replot the same RBM shifts (Figs. 3(a)) for out-of-phase and in-phase RBM oscillations as a function of both inner tube diameter as well as diameter difference between inner and outer nanotubes in Figs. 3(b) and 3(c), respectively. A clear tendency is observed for the out-of-phase oscillations in Fig. 3(b): the frequency shift increases with increasing (inner) tube diameter and with decreasing inter-tube distance. The smaller the inter-tube distance, the stronger coupling between two RBMs of inner and outer nanotubes, which explains the latter trend. How to understand the frequency shift becomes larger with increasing (inner) tube diameter? For large-diameter DWNTs, the RBM frequencies of two pristine inner and outer SWNTs are close to each other, and thus the coupling between inner and outer walls tends to be strong, giving rise to larger frequency blueshifts. This is analogous to a larger energy correction term in second-order purtaburbation if two unperturbed states are close in energy. By the same token, small-diameter DWNTs show weaker vibrational coupling and smaller frequency blueshifts, so that the vibrational eigenmodes for the in-phase and out-of-phase RBM oscillations are mainly comprised of lattice motions of constituent outer nanotube and inner nanotube, respectively. We note that a weaker effect is present for the in-phase oscillations in Fig. 3(c), which is presumably due to the more significant influence of environment effect on outer nanotube that dominates the in-phase motion.

**Supplementary Note 5**

Our DFT calculations in Fig. 4(d) quantitatively match the experimental results for two pristine RBMs of (7,7) (~239.5 vs 240.3 cm$^{-1}$) and (12,12) (~142.5 vs ~140.05 cm$^{-1}$) as well as for the coupled in-phase (~150.1 vs ~151.9 cm$^{-1}$) and out-of-phase (~253.9 vs ~252.0 cm$^{-1}$) oscillations in DWNT (7,7) @ (12,12). Given that our Raman measurement uncertainty is about 1–2 cm$^{-1}$, the agreement between our experiment and DFT calculation is remarkable. The calculated in-phase (23) and out-of-phase (36) eigenmodes also verify their weakly-coupled characters. All these results confirm the high fidelity and accuracy for both experiment and theory.

**Supplementary Figure 9.**

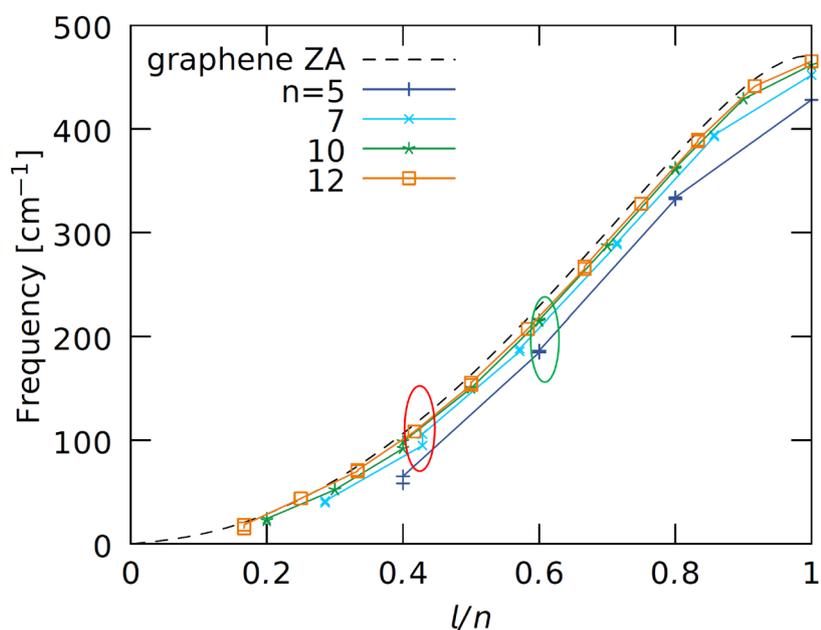

**Fig. S9. Calculated eigenvalues for out-of-plane transverse acoustic modes (ZA modes) as a function of *l/n* for armchair nanotubes (*n,n*) with *n*=5, 7, 10 and 12.** The two nearly degenerate ZA modes (*l*=3 for (7,7) and *l*=5 for (12,12)) that are responsible for the new phonon mode in (7,7) @ (12,12) are highlighted by the red circle. The curvature effect is noticeable as the ZA mode eigenvalues deviate more and more from those of 2D graphene as *n* is reduced. This suggests that curvature effect needs to be taken into account in addition to the quantum confinement (i.e., zone folding scheme) to evaluate the coupling condition for small-diameter nanotubes. The green circle highlights the curvature effect on ZA mode eigenvalues for (5,5) and (10,10), which makes the otherwise degenerate two states (*l*=3 for (5,5) and *l*=6 for (10,10)) have different energies.

**Supplementary Table 2**

| Mode | Frequency (cm$^{-1}$) | $W_{inner}$ | $W_{outer}$ | $W = W_{outer} - W_{inner}$ | Approximated cross-section deviation of the inner tube | Approximated cross-section deviation of the outer tube |
|---|---|---|---|---|---|---|
| 17 | 99.504 | 0.000 | 1.000 | 1.000 | 0.005 | -0.028 |
| 18 | 99.621 | 0.000 | 1.000 | 1.000 | 0.002 | -0.002 |
| 19 | 120.163 | 0.866 | 0.134 | -0.731 | -0.083 | -0.432 |
| 20 | 122.652 | 0.018 | 0.982 | 0.965 | 0.001 | -0.137 |
| 21 | 126.716 | 0.499 | 0.501 | 0.002 | -0.042 | -0.993 |
| 22 | 127.151 | 0.395 | 0.605 | 0.210 | -0.038 | 0.964 |
| 23 | 150.142 | 0.010 | 0.990 | 0.980 | -0.542 | -7.012 |
| 24 | 157.393 | 1.000 | 0.000 | -0.999 | 0.022 | 0.133 |
| 25 | 157.445 | 1.000 | 0.000 | -1.000 | 0.012 | 0.076 |
| 26 | 167.000 | 0.002 | 0.998 | 0.996 | 0.002 | 0.001 |
| 27 | 169.432 | 0.001 | 0.999 | 0.997 | 0.024 | 0.046 |
| 28 | 181.173 | 0.000 | 1.000 | 1.000 | 0.014 | 0.032 |
| 29 | 181.253 | 0.000 | 1.000 | 1.000 | 0.001 | 0.025 |
| 30 | 194.914 | 0.977 | 0.023 | -0.954 | -0.111 | -0.083 |
| 31 | 197.608 | 0.960 | 0.040 | -0.919 | -0.099 | -0.250 |
| 32 | 200.307 | 0.019 | 0.981 | 0.962 | 0.007 | 0.485 |
| 33 | 200.888 | 0.021 | 0.979 | 0.959 | -0.004 | -0.002 |
| 34 | 218.704 | 0.000 | 1.000 | 0.999 | 0.022 | 0.045 |
| 35 | 218.973 | 0.001 | 0.999 | 0.999 | -0.005 | 0.141 |
| 36 | 253.884 | 0.990 | 0.010 | -0.981 | -5.483 | 0.660 |

**Table S2. Calculated $W$ values and approximate cross-section deviations for all coupled modes shown in the middle row in Fig. 4(d) in the main text.** The new phonon modes 21 and 22, in-phase and out-of-phase RBM modes 23 and 36 are highlighted in yellow. $W$ is defined in the main text. In the last two columns, the cross-section deviations, which are defined as the cross-sections of the inner or outer tubes deformed by normalized eigenmode vectors with the equilibrium cross-sections subtracted, are summarized. In a very naïve picture, the Raman activity is related to difference between the electron polarizability at two extrema of vibration, and the volume change can be used as a measure for that difference. Larger absolute values in the last two columns mean larger cross-section deviations.

**Supplementary Figure 10.**

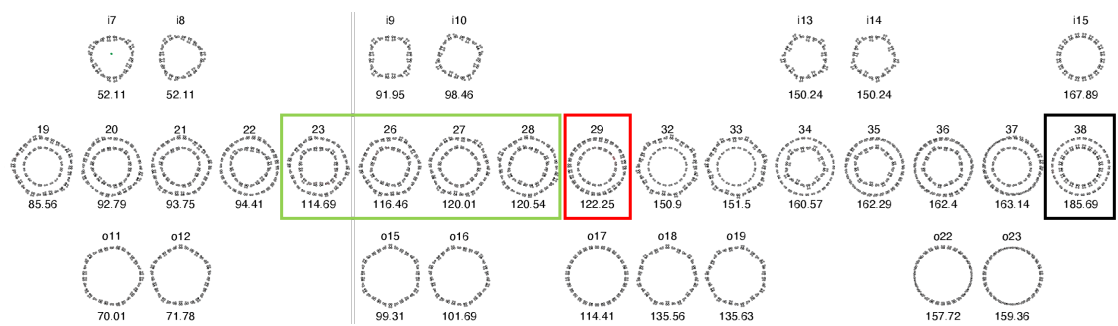

**Fig. S10. DFT calculations on the phonon eigenmodes and eigenvalues for DWNT (10,10) @ (15,15) with a diameter ratio of 2/3.** The selected phonon eigenvectors and eigenvalues in the range between 80 cm$^{-1}$ and 190 cm$^{-1}$ obtained by the calculation are shown. (The omitted modes have no radial components in their eigenvectors, indicating negligible inter-tube couplings.) The coupled vibrational eigenvectors are shown in the middle row, while those for pristine inner and outer nanotubes are shown in the top and bottom rows. The red and black solid frames mark the weakly-coupled in-phase and out-of-phase RBM modes. Green frame marks the strongly hybridized modes as a result of mixing of four ZA modes i9 and i10 (with $l$=4), o15 and o16 (with $l$=6). This observation further supports the proposed phonon coupling mechanism.

**Supplementary Figure 11.**

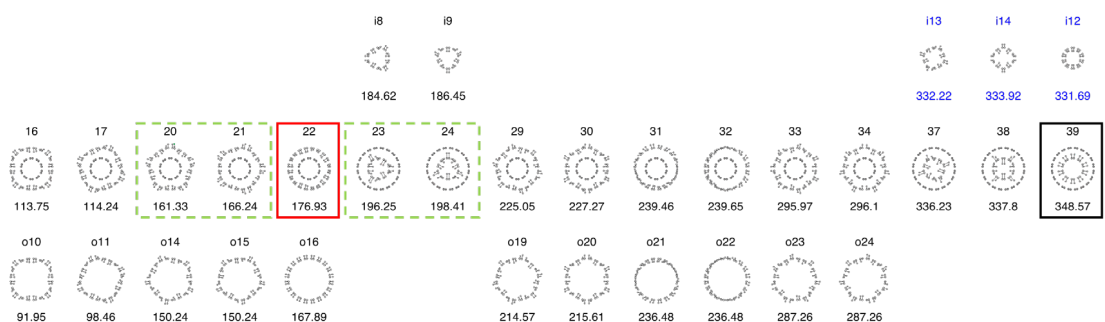

**Fig. S11. DFT calculations on the phonon eigenmodes and eigenvalues for DWNT (5,5) @ (10,10) with a diameter ratio of 1/2.** The selected phonon eigenvectors and eigenvalues in the range between 95 cm$^{-1}$ and 350 cm$^{-1}$ obtained by the calculation are shown. (The omitted modes have no radial components in their eigenvectors, indicating negligible inter-tube couplings.) The coupled vibrational eigenvectors are shown in the middle row, while those for pristine inner and outer nanotubes are shown in the top and bottom rows. The red and black solid frames mark the weakly-coupled in-phase and out-of-phase RBM modes. Green dashed lines mark the coupled modes as a result of mixing of four ZA modes among i8, i9, o14 and o15. This coupling is not strong as the motions of the four modes 20, 21, 23 and 24 are dominated by motion of either inner or outer wall. The large energy mismatch between i7/i8 and o14/o15 leads to the weak

phonon coupling, in contrast to the (7,7) @ (12,12) in the main text. We note that the curvature effect is also important to observe strong phonon-phonon coupling. For example, considering quantum confinement (i.e., zone folding) only would predict a strong coupling between i8/i9 (ZA mode with $l=3$) and o19/o20 (ZA mode with $l=6$) for the diameter ratio between inner (5,5) and outer (10,10) is $1/2$. However, the DFT results including the curvature effect show that they have very different energies (see Fig. S9).

**Supplementary Note 6.**

Suppose a one-dimensional system consists of two periods $l_1$ and $l_2$ with $l_1 < l_2$. Then, there arises a moiré period $L$, which contains $s+1$ units of $l_1$ cell and $s$ units of $l_2$ cell. Then, we have $L = (s+1)l_1 = sl_2$. From this, $s = l_1/(l_2 - l_1)$ and $L = l_1 l_2/(l_2 - l_1)$ are derived. When $s$ is an integer, the moiré period matches to the period of the underlying lattice. When $s$ is a non-integer rational number, the underlying lattice is periodic, but its period does not match to the moiré period. If $s$ is an irrational number, the underlying lattice is aperiodic.

In nanotubes, it is convenient to use azimuthal angle as a coordinate. A $(n,n)$-nanotube (i.e., armchair nanotube) has a period of $2\pi/n$ in the azimuthal angle space. For double-walled armchair nanotubes, we have $l_1 = 2\pi/n_{\text{outer}}$ and $l_2 = 2\pi/n_{\text{inner}}$, which leads to $L = 2\pi/(n_{\text{outer}} - n_{\text{inner}})$. On the other hand, for the armchair cases, $n_{\text{outer}} - n_{\text{inner}} = 5$ is the condition for realistic fabrication of a double-walled carbon nanotube, since it makes its inter-tube distance well matches to the layer distance in bilayer graphene. Then, local environments along the azimuthal direction change with the period of $2\pi/5$, which accounts the pentagonal distortions assuming that the inter-tube phononic coupling is fixed by local environments.

**References:**

[1] Ceresoli, G. L. Chiarotti, M. Cococcioni, I. Dabo *et al.*, QUANTUM ESPRESSO: a modular and open-source software project for quantum simulations of materials, Journal of Physics: Condensed Matter **21**, 395502 (2009).

[2] P. Giannozzi, O. Andreussi, T. Brumme, O. Bunau, M. Buongiorno Nardelli, M. Calandra, R. Car, C. Cavazzoni, D. Ceresoli, M. Cococcioni *et al.*, Advanced capabilities for materials modelling with Quantum ESPRESSO, Journal of Physics: Condensed Matter **29**, 465901 (2017).

[3] A. Togo, L. Chaput, T. Tadano, and I. Tanaka, Implementation strategies in phonopy and phono3py, Journal of Physics: Condensed Matter **35**, 353001 (2023).

[4] A. Togo, First-principles Phonon Calculations with Phonopy and Phono3py, Journal of the Physical Society of Japan **92**, 012001 (2022).

[5] A. Dal Corso, Pseudopotentials periodic table: From H to Pu, Computational Materials Science **95**, 337 (2014).

[6] S. Grimme, J. Antony, S. Ehrlich, and H. Krieg, A consistent and accurate ab initio parametrization of density functional dispersion correction (DFT-D) for the 94 elements H-Pu, The Journal of Chemical Physics **132**, 154104 (2010).

[7] K. Liu, Z. Xu, W. Wang, P. Gao, W. Fu, X. Bai, and E. Wang, Direct determination of atomic structure of large-indexed carbon nanotubes by electron diffraction: application to double-walled nanotubes, Journal of Physics D: Applied Physics **42**, 125412 (2009).

[8] M. Gao, J. M. Zuo, R. Zhang, and L. A. Nagahara, Structure determinations of double-wall carbon nanotubes grown by catalytic chemical vapor deposition, Journal of Materials Science **41**, 4382 (2006).

[9] M. Koshino, P. Moon, and Y.-W. Son, Incommensurate double-walled carbon nanotubes as one-dimensional moiré crystals, Physical Review B **91**, 035405 (2015).

[10] K. Liu, C. Jin, X. Hong, J. Kim, A. Zettl, E. Wang, and F. Wang, Van der Waals-coupled electronic states in incommensurate double-walled carbon nanotubes, Nature Physics **10**, 737 (2014).